\documentclass[onecolumn,aps,floatfix,pra,superscriptaddress,nofootinbib]{revtex4}
\usepackage{hyperref}
\usepackage{color}
\usepackage{graphicx}
\usepackage{bm}
\usepackage{amsmath}
\usepackage{enumerate}
\usepackage{upgreek}
\usepackage[subnum]{cases}

\newcommand{\cl}{ \text{cl} }
\newcommand{\qm}{ \text{q} }
\newcommand{\eff}{ \text{eff} }
\newcommand{\tr}{ \text{tr} }

\newcommand{\pa}{ \partial }
\newcommand{\hb}{ \hbar }
\newcommand{\si}{ \sigma }
\newcommand{\om}{ \omega }

\newcommand{\ga}{ \gamma }
\newcommand{\la}{ \langle }
\newcommand{\ra}{ \rangle }
\newcommand{\del}{ \delta }
\begin{document}
\title{On non-linear Schr\"{o}dinger equations for open quantum systems }
\author{S. V. Mousavi}
\email{vmousavi@qom.ac.ir}
\affiliation{Department of Physics, University of Qom, Ghadir Blvd., Qom 371614-6611, Iran}
\author{S. Miret-Art\'es}
\email{s.miret@iff.csic.es}
\affiliation{Instituto de F\'isica Fundamental, Consejo Superior de
Investigaciones Cient\'ificas, Serrano 123, 28006 Madrid, Spain}
\begin{abstract}

Recently two generalized nonlinear Schr\"{o}dinger equations have been proposed  by Chavanis [Eur. Phys. J. Plus 132 (2017) 286] by applying Nottale's theory of scale relativity relying on a fractal space-time to describe dissipation in quantum systems. Several existing 
nonlinear equations are then derived and discussed in this context leading to a continuity equation with an extra source/sink term which violates 
Ehrenfest theorem. 
An extension to describe stochastic dynamics is also carried out  by including thermal fluctuations or noise of the environment.
These two generalized nonlinear equations are analyzed within the Bohmian mechanics framework to describe the corresponding dissipative 
and stochastic dynamics in terms of quantum trajectories. Several applications of this second generalized equation which can be considered 
as a generalized  Kostin equation have been carried out. The first application consists of the study of the position-momentum uncertainty 
principle in a dissiaptive dynamics. 
After,  the so-called Brownian-Bohmian motion is investigated by calculating classical and quantum diffusion coefficients. 
And as a third example, transmission through a transient (time dependent) parabolic repeller is studied where the interesting
phenomenon of early arrival is observed even in the stochastic dynamics although the magnitude of early arrival is reduced by friction.

\end{abstract}
\maketitle

\section{Introduction}

Since perfect isolation of quantum systems is not possible, any realistic quantum system is influenced by its environment. By taking into account 
the interaction of the system with its environment, the corresponding dynamics becomes stochastic.
There are three main approaches in the literature to study dissipative and stochastic effects in open quantum systems \cite{NaMi-book-2017}:
(i) the system-plus-environment approach leading to master equations describing the reduced density matrix for the system, (ii) explicitly 
time-dependent Hamiltonians, simple but problematic since, for example, the Heisenberg uncertainty principle is violated and 
(iii) non-linear Schr\"{o}dinger equations.  Seven of such non-linear Schr\"{o}dinger equations have been analyzed in \cite{BaNa-JAMA-2012} 
by providing the corresponding Feynman propagators: the Bialynicki-Birula and Mycielski (BBM) equation \cite{BBMy-AP-1976}, 
the Bateman-Caldirola-Kanai equation \cite{Bateman,Caldirola,Kanai}, the Di\'osi-Halliwell-Nassar equation \cite{Diosi,BaNa-JAMA-2012}, 
the Kostin equation \cite{Ko-JCP-1972}, the Schuch-Chung-Hartmann (SCH) equation \cite{ScChHa-JMP-1983, Sc-IJQC-1999}, 
the S\"ussmann-Hasse-Albretch-Kostin-Nassar equation \cite{Sussmann,Hasse,Albrecht,Kostin,Nassar1} and  
the Schr\"odinger-Nassar equation \cite{Nassar2}.
Among these equations, the most popular and important one is maybe the so-called Schr\"{o}dinger-Langevin equation or Kostin equation which 
was derived heuristically by Kostin  from the Heisenberg-Langevin equation for the momentum operator. 

Recently, a non-linear Schr\"{o}dinger equation was also proposed by Nassar and Miret-Art\'es (NM) \cite{NaMi-PRL-2013} to describe the continuous measurement of the position of a quantum particle interacting with its environment. More recently, Chavanis \cite{Ch-EPJP-2017} derived two other non-linear equations using the theory of scale relativity \cite{No-book-2011}. In this way, he generalized the Nottale's approach by including a damping force in the fundamental equation of dynamics. If a friction force $-\ga \mathbf{U}$ is naively introduced in the scale-covariant  equation of dynamics then a damped generalized Schr\"{o}dinger equation is obtained that violates local conservation of the normalization condition.  Thus, in order to obtain an equation which maintains local conservation of probability density, he took into account the friction via introducing Re$(-\ga \mathbf{U})$ in Newton's law of motion.
When written in polar form the wave function, the former generalized Schr\"{o}dinger equation reduces to the Schuch-Chung-Hartman equation  for real friction coefficient. But, one should note that an additional condition supplies the SCH equation. 
It is seen that the SCH equation is a special case of the NM equation in their polar forms.
The later generalized Schr\"{o}dinger equation reduces to BBM equation \cite{BBMy-AP-1976} for imaginary friction coefficient while it is 
equivalent to the Kostin equation \cite{Ko-JCP-1972}, without noise and for a real friction coefficient. A different generalized Schr\"odinger-Langevin
equation has been proposed in the literature for nonlinear interaction providing a state-dependent dissipation process exhibiting multiplicative noise
\cite{BaMi-AOP-2014}. This equation was after extended to a non-Markovian problem \cite{VaMoBa-AOP-2015}.

In this work, the two proposed generalized  nonlinear equations by Chavanis have been extended to describe stochastic dynamics by including 
thermal fluctuations or noise of the environment. Both generalized nonlinear equations are analyzed within the context of Bohmian 
mechanics to describe the corresponding dissipative and stochastic dynamics in terms of quantum trajectories. 
The first generalized Schr\"{o}dinger equation is also shown to violate Ehrenfest theorem as well as the SCH and NM equations 
due to the fact that they lead to a continuity equation with an extra source/sink term. On the contrary,
the second generalized Schr\"{o}dinger equation preserves local conservation of probability density. Finally, several applications of this 
second generalized equation which can be considered as a generalized  Kostin equation 
have been carried out. The first application consists of the study of the position-momentum uncertainty principle in a dissiaptive dynamics. 
After,  the so-called Brownian-Bohmian motion is investigated by calculating classical and quantum diffusion coefficients. 
And as a third example, transmission through a transient (time dependent) parabolic repeller is studied where the interesting
phenomenon of early arrival is observed even in the stochastic dynamics although the magnitude of early arrival is reduced by friction.


\section{Derivation of the generalized Schr\"{o}dinger equation}

In this section we summarise Chavanis' work \cite{Ch-EPJP-2017} in deriving a generalized Schr\"{o}dinger equation for open quantum systems and extend its range of applicability to take into account stochasticity in addition to dissipation by including the random force $ \mathbf{F}_r(t) $ in the fundamental equation of motion. The classical Langevin equation for a particle in three dimensions reads as
\begin{eqnarray} \label{eq: La_cl}
\frac{d \mathbf{u} }{dt} &=& - \frac{1}{m} \nabla ( V + V_r) - \ga \mathbf{u}
\end{eqnarray}
where $\mathbf{u}$ is the velocity of the particle, $m$  its mass, $ V $ is the interaction potential and 
\begin{eqnarray} \label{eq: ran_pot}
V_r(\mathbf{r}, t) &=& \mathbf{r} \cdot \mathbf{F}_r(t)
\end{eqnarray}
being the random potential which is linear with the position. The random force $\mathbf{F}_r(t)$ is a time dependent function and $\ga$ is a
friction coefficient which is assumed to be real.

Based on  Nottale's theory \cite{Ch-EPJP-2017, No-book-2011}, the quantum equation of motion can be derived by replacing the standard 
velocity $\mathbf{u}$ by the complex velocity $\mathbf{U}$ and the standard time derivative $d/dt$ by the complex one according to
\begin{eqnarray} \label{eq: com_der}
\frac{D}{D t} &=& \frac{\pa}{\pa t} + \mathbf{U} \cdot \nabla - i \mathcal{D}  \nabla^2
\end{eqnarray}
with $ \mathcal{D} $ playing the role of a diffusion coefficient which will be determined later on.
Following Chavanis and taking eq. (\ref{eq: La_cl}) as the fundamental equation of motion, according to 
Nottale's approach we have
\begin{eqnarray} \label{eq: fun_dy}
\frac{D \mathbf{U}}{D t} &=& - \frac{1}{m}\nabla ( V + V_r ) - \ga \mathbf{U} 
\end{eqnarray}
where the friction coefficient $ \ga $ is taken to be a complex quantity  \cite{Ch-EPJP-2017}. 
Then, the complex impulse $\mathbf{P} = m\mathbf{U}$ and the complex energy  
\begin{eqnarray} \label{eq: impuls&en}
\mathbf{P} &=& \nabla \mathcal{S}, \qquad \mathcal{E} = - \frac{ \pa \mathcal{S} }{ \pa t }
\end{eqnarray}
are introduced in the Lagrangian formalism where $ \mathcal{S}(\mathbf{r}, t) $ is the complex action.
The complex velocity field is then 
\begin{eqnarray} \label{eq: complexU}
\mathbf{U} &=& \frac{\nabla \mathcal{S} }{m}   .
\end{eqnarray}
By using eqs. (\ref{eq: com_der}) and (\ref{eq: complexU}), eq. (\ref{eq: fun_dy}) can be recast as
\begin{eqnarray} \label{eq: der_HJ}
\nabla \left \{ \frac{ \pa \mathcal{S} }{ \pa t } + \frac{1}{2m} (\nabla \mathcal{S})^2 - i \mathcal{D} \nabla^2 \mathcal{S} + V + V_r + \ga \mathcal{S} \right\} &=& 0
\end{eqnarray}
which after integrating over space, the Hamilton-Jacobi equation
\begin{eqnarray} \label{eq: HJ}
\frac{ \pa \mathcal{S} }{ \pa t } + \frac{1}{2m} (\nabla \mathcal{S})^2 - i \mathcal{D} \nabla^2 \mathcal{S} + V + V_r + \ga \mathcal{S} + f(t) &=& 0
\end{eqnarray}
is obtained where $f(t)$ is the constant of integration which will be determined later on. By introducing now the wave function as follows
\begin{eqnarray} \label{eq: wf}
\mathcal{S} &=& -2 i m \mathcal{D} \ln \psi ,
\end{eqnarray}
eq. (\ref{eq: HJ}) can be rewritten as
\begin{eqnarray} \label{eq: eq1}
2 i m \mathcal{D} \frac{\pa \psi}{\pa t} &=& ( -2 m \mathcal{D}^2 \nabla^2  + V + V_r - 2 i m \mathcal{D} \ga \ln \psi + f(t) ) \psi   
\end{eqnarray}
and defining the diffusion coefficient after Nelson \cite{Ne-PR-1966} as 
\begin{eqnarray} \label{eq: D}
\mathcal{D} &=&  \frac{\hb}{2m}
\end{eqnarray}
eq. (\ref{eq: eq1}) leads to a generalized nonlinear Schr\"odinger equation 
\begin{eqnarray} \label{eq: gen_SCH1}
i \hb \frac{\pa}{\pa t} \psi(\mathbf{r}, t) &=& \left\{
-\frac{\hb^2}{2m} \nabla^2 + V + V_r - i \ga^{}_R \hb ( \ln\psi - \la \ln\psi \ra )+ \hb \ga^{}_I ( \ln\psi - \la \ln\psi \ra ) \right\} \psi(\mathbf{r}, t)
\end{eqnarray}
where $\ga^{}_R$ and $ \ga^{}_I $ are the real and imaginary parts of the complex friction $\ga$ playing the role of {\it classical} and 
{\it quantum} friction coefficients, respectively. The imaginary part has been related to an effective temperature $T_e$ (positive or negative) through 
$2 k_B T_e / \hbar$ where $k_B$ is the Boltzmann constant \cite{Ch-EPJP-2017}. When considering an ensemble of particles, the symmetry of the 
wave function is responsible for the appearance of what is called statistical potential \cite{Pathria} which can be attractive or repulsive. 
The term accompanying $\gamma_I$ in eq. (\ref{eq: gen_SCH1}) can then be considered as a statistical potential  leading to under certain 
conditions  the Gross-Pitaevskii equation \cite{Ch-EPJP-2017}. The integration constant $f(t)$ has been set in such a way that the expectation 
value of the friction term proportional to $\ga^{}_R$ is zero.

If the wave function is now expressed in polar form as
\begin{eqnarray} \label{eq: wf_polar}
\psi(\mathbf{r}, t) &=& |\psi(\mathbf{r}, t)| e^{iS(\mathbf{r}, t)/\hb}
\end{eqnarray}
and substituted into eq. (\ref{eq: gen_SCH1}) then the real and imaginary parts of the resulting equations yield  to
\begin{numcases}~ 
-\frac{ \pa S }{ \pa t } = \frac{1}{2m} (\nabla S)^2 + V + V_r + Q + \ga^{}_R (S - \la S \ra) + \frac{\hb \ga^{}_I}{2} ( \ln \rho - \la \ln \rho \ra ) \label{eq: QHJ1}
\\
\frac{ \pa \rho }{ \pa t } + \nabla \cdot \left( \rho \frac{\nabla S}{m} \right) = - \ga^{}_R \rho ( \ln \rho - \la \ln \rho \ra ) + \frac{2 \ga^{}_I }{\hb} \rho ( S - \la S \ra )
\label{eq: con1}
\end{numcases}
respectively, where $ \rho = |\psi|^2$ is the probability density and
\begin{eqnarray} \label{eq: Qp}
Q(\mathbf{r}, t) &=& - \frac{\hb^2}{2m} \frac{ \nabla^2 |\psi| }{ |\psi| }  
\end{eqnarray}
is the so-called quantum potential. Eqs. (\ref{eq: QHJ1}) and (\ref{eq: con1}) can be seen as the generalized Hamilton-Jacobi and continuity equations.
The corresponding continuity equation with two source/sink terms clearly shows that eq. (\ref{eq: gen_SCH1}) violates the local conservation 
of probability density. However, the integration over the whole space of eq. (\ref{eq: con1}) reveals the correct global conservation of the normalization.
The partial derivative  with respect to the space coordinate  of eq. (\ref{eq: QHJ1}) yields 
\begin{eqnarray} \label{eq: nabla_QHJ1}
\left( \frac{ \pa }{ \pa t } + \frac{1}{m} \nabla S \cdot \nabla \right) \nabla S
&=& - \nabla \left( V + V_r + Q  \right)  - \ga^{}_R \nabla S - \frac{\hb \ga^{}_I}{2 \rho} \nabla \rho
\end{eqnarray}

\subsection{Gaussian wave packet dynamics}

For simplicity, we now consider the one-dimensional dissipative motion, $V_r =0$, and solve eq. (\ref{eq: gen_SCH1}) or equivalently 
eqs. (\ref{eq: con1}) and (\ref{eq: nabla_QHJ1}) by means of a time-dependent Gaussian ansatz for the probability density \cite{NaMi-book-2017}
\begin{eqnarray} \label{eq: rho_ansatz}
\rho(x, t) &=& \frac{1}{\sqrt{2\pi}\si(t)} \exp \left[ -\frac{(x-q(t))^2}{2 \si^2(t)} \right] ,
\end{eqnarray}
where $q(t) = \int dx ~ x \rho(x, t) $ is the expectation value of the position operator giving the center of the corresponding Gaussian 
wave packet, $\si(t)$ being its width. 
For this goal, we first divide eq. (\ref{eq: con1}) by $\rho$ and then derive the resulting equation with respect to $x$ to have just some derivatives 
of $S$. In this way, we obtain
\begin{eqnarray} \label{eq: aux1}
\frac{\pa^3 S}{\pa x^3} + \frac{1}{\rho} \frac{\pa \rho}{\pa x} \frac{\pa^2 S}{\pa x^2} + 
\left[ - \frac{2m\ga_I}{\hb} - \frac{1}{\rho^2} \left(\frac{\pa \rho}{\pa x} \right)^2 + \frac{1}{\rho} \frac{\pa^2 \rho}{\pa x^2} \right] \frac{\pa S}{\pa x}
+ \frac{m}{\rho} \left( \ga^{}_R \frac{\pa \rho}{\pa x} - \frac{1}{\rho} \frac{\pa \rho}{\pa x} \frac{\pa \rho}{\pa t} + \frac{\pa^2 \rho}{\pa x \pa t}
\right) &=& 0    .
\end{eqnarray}
This equation can be solved by the linear (in space) ansatz
\begin{eqnarray} \label{eq: pa_xS ansatz}
\frac{\pa S}{\pa x} &=& a(t) ( x - q(t) ) + b(t)
\end{eqnarray}
and then by introducing ansatzs (\ref{eq: rho_ansatz}) and (\ref{eq: pa_xS ansatz}) into eq. (\ref{eq: aux1}) the corresponding time dependent 
coefficients $a(t)$ and $b(t)$ are given by
\begin{eqnarray} 
a(t) &=& m \frac{ \dot{\si}(t) - \frac{\ga^{}_R}{2} \si(t) }{\si(t) \left( 1 + \frac{m \ga^{}_I}{\hb}\si(t)^2 \right) } \label{eq: at} \\
b(t) &=& m \frac{ \dot{q}(t)}{ 1 + \frac{2 m \ga^{}_I}{\hb}\si(t)^2 }  \label{eq: bt}
\end{eqnarray}
where the linear independence of different powers of $ x - q(t) $ has been used.
Thus, we have that
\begin{eqnarray} \label{eq: S}
S(x, t) &=& \frac{m}{2} \frac{ \dot{\si}(t) - \frac{\ga^{}_R}{2} \si(t) }{\si(t) \left( 1 + \frac{m \ga^{}_I}{\hb}\si(t)^2 \right) } ( x - q(t) )^2 + m \frac{ \dot{q}(t)}{ 1 + \frac{2 m \ga^{}_I}{\hb}\si(t)^2 } ( x - q(t) ) + g(t)
\end{eqnarray}
where $g(t)$ is the constant of integration which can be determined by eq. (\ref{eq: QHJ1}), its explicit form  being not important for us since 
$S$ always appears together with its average as $ S - \la S \ra $. The center of the wave packet $q(t)$ and its width $\si(t)$ are obtained 
by introducing eq. (\ref{eq: S}) into eq. (\ref{eq: nabla_QHJ1}) and using the wave packet approximation and expanding the interaction 
potential $V(x)$ around $q(t)$ up to second order we have that
\begin{equation} \label{eq: sigma(t)}
\begin{aligned}
 & \left( 1 + \frac{m \ga^{}_I \si(t)^2}{\hb} \right) \ddot{\si}(t)
- \frac{3 m \ga^{}_I \si(t)}{\hb} \dot{\si}(t)^2
+ \frac{2 m \ga^{}_I \ga^{}_R \si(t)^2}{\hb}\dot{\si}(t)
+ \frac{m V_2 \ga_I^2}{\hb^2}\si(t)^5  \\
      & + \frac{\ga^{}_I}{2 \hb} \left( 4 V_2 - m \ga_I^2 - m \ga_R^2 \right) \si(t)^3
+ \left(  \frac{V_2 }{m} - \frac{\ga_R^2 + 5 \ga_I^2}{4} \right) \si(t)
- \frac{\ga^{}_I \hb}{m \si(t)} - \frac{\hb^2}{4 m^2 \si(t)^3} = 0
\end{aligned}
\end{equation}
and
\begin{equation} \label{eq: q(t)}
\begin{aligned}
 & \left[ 1 + \ga^{}_I \left( \frac{3 m \si(t)^2}{\hb} + \frac{2 m^2 \ga^{}_I \si(t)^4}{\hb^2} \right) \right] \ddot{q}(t) \\
      & + \left[ \ga^{}_R + 
\ga^{}_I \left( \frac{4 m  \ga^{}_R \si(t)^2}{\hb} + \frac{4 m^2 \ga^{}_I \ga^{}_R \si(t)^4}{\hb^2} 
- \frac{6 m \si(t) \dot{\si}(t)}{\hb} - \frac{8 m^2 \ga^{}_I \si(t)^3 \dot{\si}(t)}{\hb^2} 
\right)
\right] \dot{q}(t) \\
      & = - \frac{V_1}{m} \left[ 1 + \ga^{}_I \left(\frac{5 m \si(t)^2}{\hb} + \frac{8 m^2 \ga^{}_I \si(t)^4}{\hb^2} 
+ \frac{4 m^3 \ga_I^2 \si(t)^6}{\hb^3}
\right)\right]
\end{aligned}
\end{equation}
where
\begin{eqnarray} 
V_1 &=& \frac{\pa V}{\pa x} \bigg|_{x=q(t)} ~ \mbox{and} ~~~~
V_2 = \frac{\pa^2 V}{\pa x^2} \bigg|_{x=q(t)}
\end{eqnarray}
and the linear independency of different powers of $ x - q(t) $ has been again used.
Within this approximation which is exact for potentials of at most second order in space coordinates, one should expect that the center 
of the wave packet follows a classical trajectory. However, eq. (\ref{eq: q(t)}) explicitly shows that $q(t)$ does not follow  a classical trajectory.
The wave property of the particle (its width) is involved in the equation of motion for the center of the wave packet. On the other hand, this 
width is only ruled by eq. (\ref{eq: sigma(t)}). For the special case of a  real friction coefficient, $q(t)$ follows the classical equation of motion 
and the differential equation for the width is the well-known dissipative Pinney equation \cite{NaMi-book-2017}. 


\subsection{Ehrenfest relations}

We have seen in the previous section that Ehrenfert's theorem is not fulfilled for complex frictions.
As is well known, Ehrenfest's relations are given by 
\begin{numcases}~ 
\frac{d}{dt} \la \hat{ \mathbf{r} } \ra = \frac{ \la \hat{ \mathbf{p} } \ra }{m} \label{eq: Eh1} 
\\
\frac{d}{dt} \la \hat{ \mathbf{p} } \ra = - \la \nabla V \ra - \ga^{}_R \la \hat{ \mathbf{p} } \ra \label{eq: Eh2}
\end{numcases}
and required for the correspondence principle.
For interaction potentials of at most second order in space coordinates where $ \la \nabla V(\mathbf{r}) \ra = \nabla V(\la \mathbf{r} \ra) $, 
after eq. (\ref{eq: Eh1}), the second Ehrenfest relation (\ref{eq: Eh2}) can be written as 
\begin{eqnarray}\label{eq: Eh_theor}
m \frac{d^2}{dt^2} \la \hat{ \mathbf{r} } \ra &=& - \nabla V(\la \mathbf{r} \ra) - \ga^{}_R \la \hat{ \mathbf{p} } \ra
\end{eqnarray} 
which is just the classical equation of motion for the expectation value of the position operator. This result is known as the Ehrenfest theorem
\cite{BaYaZi-PRA-1994}. One can see that the standard continuity equation i.e., the continuity equation without source/sink terms, which preserves 
local conservation of normalization, is a {\it sufficient} condition for fulfilment of the Ehrenfest relation (\ref{eq: Eh1}).

We now show that the generalized Schr\"{o}dinger equation (\ref{eq: gen_SCH1}) violates both Ehrenfest relations in general.
The time-derivative of the expectation value of the position operator is given by
\begin{eqnarray} \label{eq: d<r>_dt}
\frac{d}{dt} \la \hat{ \mathbf{r} } \ra &=& \int d^3x  \frac{\pa \rho}{\pa t} \mathbf{r}
= \frac{ \la \hat{ \mathbf{p} } \ra }{m} - \ga^{}_R \big \la \hat{ \mathbf{r} } ( \ln \rho - \la \ln \rho \ra )  \big \ra
+ \frac{2\ga^{}_I}{\hb} \big \la \hat{ \mathbf{r} } ( S - \la S \ra )  \big \ra 
\end{eqnarray}
where we have used eq. (\ref{eq: con1}), the technique of integration by parts and  the relation $ \la \hat{ \mathbf{p} } \ra = \la \nabla S \ra $. 
Eq. (\ref{eq: d<r>_dt}) explicitly shows violation of (\ref{eq: Eh1}). 
For the Gaussian solution (\ref{eq: rho_ansatz}) where $S$ is given by eq. (\ref{eq: pa_xS ansatz}), eq. (\ref{eq: d<r>_dt}) can be used to deduce the 
coefficient $b(t)$
%
from which eq. (\ref{eq: bt}) is obtained.

On the other hand, the second Ehrenfest relation (\ref{eq: Eh2}) is also violated. To this end, we multiply both sides of eq. (\ref{eq: nabla_QHJ1}) 
by $\rho$. Then by taking into account the continuity equation (\ref{eq: con1}) we obtain
\begin{equation} \label{eq: aux2}
\begin{aligned}
 & \frac{\pa ( \rho \nabla S )}{\pa t} + \sum_i \mathbf{e}_i \nabla \cdot \left( \rho \frac{\nabla_i S}{m} \nabla S \right) + \nabla S \left( \ga^{}_R \rho( \ln\rho - \la \ln\rho \ra) - \frac{2\ga_I}{\hb} \rho ( S - \la S \ra )   \right) 
 \\ 
 & = - \rho \nabla(V + Q) - \ga^{}_R \rho \nabla S - \frac{\hb \ga^{}_I}{2} \nabla \rho
\end{aligned}
\end{equation}
where $ \mathbf{e}_i $ denotes the unit vector along $x_i$ direction .
Now by integrating both sides of this equation over the whole space and noting that $ \la \nabla Q \ra = 0 $, it yields
\begin{eqnarray} \label{eq: ehren}
\frac{d}{d t} \la \nabla S \ra &=& - \la \nabla V \ra 
- \ga^{}_R \la \nabla S \ra
- \ga^{}_R \bigg \la \nabla S ( \ln\rho - \la \ln\rho \ra )  \bigg \ra
+ \frac{2\ga^{}_I}{\hb} \bigg \la  \nabla S ( S - \la S \ra )  \bigg \ra
\end{eqnarray}
%
For solutions such that the phase of the wave function is linear in space, $\nabla S$ is only a function of time. Thus, the last two terms of RHS of 
eq. (\ref{eq: ehren}) becomes zero. In such a case, Ehrenfest relation is fulfilled.
Note that for the Gaussian solution (\ref{eq: rho_ansatz}) where $S$ is given by (\ref{eq: S}), $\nabla S$ depends linearly on the space coordinate. 
In this case, the middle term in RHS of eq. (\ref{eq: ehren}) is zero but not the last term, $ \la  \nabla S ( S - \la S \ra ) \ra = 2 a(t) b(t) \si(t)^2$
(it is worth mentioning that in all of the above proofs we have repeatedly used the technique of integration by parts  and set all resultant boundary 
terms equal to zero).

\subsection{The SCH and NM equations }

Schuch, Chung and Hartman (SCH) proposed the logarithmic non-linear equation \cite{ScChHa-JMP-1983, Sc-IJQC-1999} 
\begin{eqnarray}  \label{eq: SCH}
i \hb \frac{\partial}{\partial t}\psi(\mathbf{r}, t) &=& \left[ -\frac{\hb^2}{2m} \nabla^2
+ V(\mathbf{r}, t) - i \hb \ga  \left( \ln \psi - \langle \ln \psi \rangle \right) \right] \psi(\mathbf{r}, t)
\end{eqnarray}
with real $ \ga $ for the description of frictional effects in dissipative systems with the additional condition
\begin{eqnarray} \label{eq: con_on_source}
- \ga \rho ( \ln \rho - \langle \ln \rho \rangle ) &=& D(t) \nabla^2 \rho
\end{eqnarray}
with $ D(t) $ playing the role of a time-dependent diffusion coefficient. It is seen that this restricting condition 
is automatically fulfilled for the Gaussian solution (\ref{eq: rho_ansatz}).
The interesting point about eq. (\ref{eq: gen_SCH1}) is that for $ \ga^{}_I = 0 $, the SCH equation (\ref{eq: SCH}) is recovered. 
By replacing $ \ga^{}_R $ by $ \ga $, eqs. (\ref{eq: d<r>_dt}) and (\ref{eq: ehren}) take the form
\begin{numcases}~ 
\frac{d}{dt} \la \hat{ \mathbf{r} } \ra 
= \frac{ \la \hat{ \mathbf{p} } \ra }{m} - \ga \big \la \hat{ \mathbf{r} } ( \ln \rho - \la \ln \rho \ra )  \big \ra \label{eq: SCH_eh1} 
\\
\frac{d}{d t} \la \hat{ \mathbf{p} } \ra = - \la \nabla V \ra 
- \ga \la \nabla S \ra - \ga \la \nabla S ( \ln\rho - \la \ln\rho \ra ) \ra \label{eq: SCH_eh2}
\end{numcases}
in the framework of the SCH equation.
According to eq. (\ref{eq: con_on_source}), the second term of the RHS of eq. (\ref{eq: SCH_eh1}) vanishes which demonstrates 
the fulfilment of the first Ehrenfest relation by the SCH equation. Although, the last term of eq. (\ref{eq: SCH_eh2}) is zero for the Gaussian 
solution, it is not generally zero revealing the violation of the second Ehrenfest relation by the SCH equation.

The SCH equation is a special case of the more general nonlinear Schr\"odinger equation \cite{NaMi-PRL-2013}
\begin{eqnarray}  \label{eq: NM}
i \hb \frac{\pa}{\pa t}\psi(\mathbf{r}, t) &=& \left[ -\frac{\hb^2}{2m} \nabla^2
+ V + V_r(\mathbf{r}, t) +  i \hb ( W_{\text{c}}(\mathbf{r}, t) + W_{\text{f}}(\mathbf{r}, t) )
\right] \psi(\mathbf{r}, t) 
\end{eqnarray}
with
\begin{eqnarray}  
W_{\text{c}}(\mathbf{r}, t) &=& -\kappa ( \ln | \psi |^2 - \langle \ln | \psi |^2  \rangle ) \label{eq: pot_Wc} \\
W_{\text{f}}(\mathbf{r}, t) &=& -\frac{\ga}{2} \left(  \ln \frac{\psi}{\psi^*} - \left\langle  \ln \frac{\psi}{\psi^*} \right\rangle \right) \label{eq: pot_Wf}
\end{eqnarray}
with real $\ga$ and $\kappa$ proposed by Nassar and Miret-Art\'es (NM) to describe continuous measurements. 
Here, $ \kappa $ plays the role of the resolution of the continuous measurement. For the special case $ \kappa = \ga/ 2 $, it reduces to the SCH 
equation. If the polar form of the wave function (\ref{eq: wf_polar}) is substituted into eq. (\ref{eq: NM}), the resulting equations for the real 
and imaginary parts are expressed as
\begin{numcases}~ 
-\frac{\pa S}{\pa t} = \frac{1}{2m} (\nabla S)^2 + V + V_r(\mathbf{r},t) + Q + \ga ( S - \langle S \rangle ) \label{eq: HJ_NM} 
\\
\frac{\pa \rho }{\pa t} + \nabla \cdot \left( \rho \frac{\nabla S}{m} \right) = - 2 \kappa \rho ( \ln \rho - \langle \ln \rho \rangle ) \label{eq: con_NM}
\end{numcases}
where $ Q $ is defined by eq. (\ref{eq: Qp}). 
Comparison of eq. (\ref{eq: QHJ1}) with eq. (\ref{eq: HJ_NM}) and eq. (\ref{eq: con1}) with eq. (\ref{eq: con_NM}) reveals that when 
$\ga^{}_I = 0 $ and $ \ga^{}_R = 2 \kappa $, the generalized Schr\"{o}dinger equation (\ref{eq: gen_SCH1}) is equivalent to the NM 
equation (\ref{eq: NM}). 
However, it should be noted that the NM equation is more general than eq. (\ref{eq: gen_SCH1}) with a real friction coefficient since it also
takes into consideration the continuous measurement process.

For the Gaussian wave packet (\ref{eq: rho_ansatz}) which is valid for simple potentials, the RHS of eq. (\ref{eq: con_NM}) reduces to 
\begin{eqnarray}
- 2 \kappa \rho ( \ln \rho - \langle \ln \rho \rangle ) &=& 
\kappa \si^2(t) \nabla^2 \rho
\end{eqnarray}
and the probability density conserves locally with the velocity field
\begin{eqnarray} \label{eq: NMvel}
\mathbf{v}(\mathbf{r}, t) &=& \frac{\nabla S}{m} - \kappa \si^2(t) \frac{1}{\rho} \nabla \rho  .
\end{eqnarray}
With this field as the Bohmian velocity field, the Bohmian trajectory approach \cite{Holland-book-1993} coincides with those of standard 
quantum mechanics. On the contrary, if the continuity equation with a source/sink term is used for the standard velocity field of Bohmian
mechanics , $\nabla S/m$, the well known quantum equivariance property \cite{DuGoZa-JSP-1992} is not fulfilled.

\section{A generalized equation fulfilling local conservation of probability density}

In the previous section we showed that the generalized Schr\"{o}dinger equation (\ref{eq: gen_SCH1}) violates local conservation of the 
probability density function. Thus, following Chavanis \cite{Ch-EPJP-2017} we now include the friction force as Re$(-\ga \mathbf{U})$ instead 
of $-\ga \mathbf{U}$ in eq. (\ref{eq: fun_dy}). Following the same steps as previously, the new generalized equation is given by  
\begin{eqnarray} \label{eq: gen_SCH2}
i \hb \frac{\pa \psi}{\pa t} &=& \left\{
-\frac{\hb^2}{2m} \nabla^2 + V + V_r + \hb \ga^{}_I \ln(|\psi|) + \frac{\hb}{2i} \ga^{}_R \left[ \ln \left( \frac{\psi}{\psi^*} \right)  - \left \la \ln \left( \frac{\psi}{\psi^*} \right)  \right \ra  \right] \right\} \psi     .
\end{eqnarray}
It is worth mentioning that Chavanis \cite{Ch-EPJP-2017} introduced the new notation $\ga^{}_I = 2 k_B T_{\eff} /\hb$ to relate the imaginary 
part of the friction coefficient to an effective temperature and found a form of fluctuation-dissipation theorem between real and imaginary 
parts of the friction coefficient.
For real frictions, $ \ga^{}_I=0 $, eq. (\ref{eq: gen_SCH2}) reduces to the well known Schr\"{o}dinger-Langevin equation derived by 
Kostin \cite{Ko-JCP-1972}. But, for an imaginary friction coefficient, it reduces to Bialynicki-Birula and Mycielski equation 
\cite{BBMy-AP-1976} which possesses soliton-like solutions of Gaussian shape. Because of this property, eq. (\ref{eq: gen_SCH2}) could also 
be proposed to describe continuous measurements in dissipative media as an alternative to the NM equation.

\subsection{Bohmian formulation}

By using the polar form of the wave function (\ref{eq: wf_polar}) in eq. (\ref{eq: gen_SCH2}), it yields to
\begin{numcases}~ 
-\frac{\pa S }{ \pa t } = \frac{1}{2m} (\nabla S)^2 + V + V_r + \ga^{}_R (S - \la S \ra) - \frac{\hb^2}{2m}
\frac{ \nabla^2 |\psi|  }{ |\psi|  } + \hb \ga^{}_I \ln |\psi| 
\label{eq: QHJ2}
\\
\frac{ \pa \rho }{ \pa t } + \nabla \cdot \left( \rho \frac{\nabla S}{m} \right) = 0   .
\label{eq: con2}
\end{numcases}
These equations are the modified Hamilton-Jacobi and continuity equations, respectively. The space derivative of eq. (\ref{eq: QHJ2}) yields
\begin{eqnarray} \label{eq: der_QHJ2}
\left( \frac{ \pa }{ \pa t } + \frac{1}{m} \nabla S \cdot \nabla \right) \nabla S
&=& - \nabla \left( V + V_r - \frac{\hb^2}{2m}
\frac{ \nabla^2 |\psi|  }{ |\psi| } + \hb \ga^{}_I \ln |\psi| \right)  - \ga^{}_R \nabla S
\end{eqnarray}
and comparison with the classical equation of motion suggests us to define a (Bohmian) velocity field as \cite{Holland-book-1993}  
\begin{eqnarray} \label{eq: BMvel2}
\mathbf{v}(\mathbf{r}, t) &=& \frac{\nabla S}{m}
\end{eqnarray}
from which (\ref{eq: der_QHJ2}) recasts
\begin{eqnarray} \label{eq: motion2}
\frac{ d \mathbf{v} }{ d t } &=& - \frac{1}{m} \nabla(V + Q) - \ga^{}_R \mathbf{v} + \frac{1}{m} \mathbf{F}_r(t) - \frac{\hb \ga^{}_I}{2m} \nabla \ln \rho 
\end{eqnarray}
where $Q$ is again the quantum potential (\ref{eq: Qp}). One should note that the new term $ \frac{\hb \ga^{}_I}{2} \ln \rho $ is actually an 
additional contribution to the usual quantum potential $Q$. It is assumed that Bohmian particles are distributed according to the Born rule. 
Thus, Bohmian results for average values of physical quantities are just quantum expectation values.

The energy of the Bohmian particle is defined as
\begin{eqnarray} \label{eq: En2}
E(\mathbf{r}, t) &=& \frac{1}{2m} (\nabla S)^2 + V + V_r + Q + \frac{\hb \ga^{}_I}{2} \ln \rho = -\frac{ \pa S }{ \pa t } - \ga^{}_R (S - \la S \ra)  
\end{eqnarray}
where in the second equality eq. (\ref{eq: QHJ2}) has been used. Thus, its average is
\begin{eqnarray} \label{eq: expEn2}
\la E \ra &=& \int d^3 x ~ |\psi(\mathbf{r}, t)|^2 E(\mathbf{r}, t) \\
&=& \frac{1}{2m} \int d^3 x ~ ( |\psi|^2 (\nabla S)^2 + \hb^2 (\nabla |\psi|)^2) +  \int d^3x ~ |\psi|^2 (V + V_r + \hb \ga^{}_I \ln |\psi|) \\
&=& \frac{\la \hat{\mathbf{p}}^2 \ra }{2m} + \la V \ra + \la V_r \ra + \frac{\hb \ga^{}_I}{2} \la \ln \rho \ra    .
\end{eqnarray}
As a consistency check of this result, we note that the energy expectation value calculated by \cite{Ko-JCP-1972} 
\begin{eqnarray} \label{eq: expEn2}
\la E \ra &=& \int d^3x ~ \psi^* i\hb \frac{\pa}{\pa t} \psi
\end{eqnarray}
leads to the same result. The rate of change of the energy expectation value with time can be known if we recall that for any arbitrary function
$A(\mathbf{r}, t)$ we have,
\begin{eqnarray} \label{eq: property}
\frac{d}{dt} \la A \ra &=& \int d^3x \frac{\pa}{\pa t} (A |\psi|^2) =  \left \la \frac{dA}{dt} \right\ra
\end{eqnarray}
where in the second equality eq. (\ref{eq: con2}) and an integration by parts have been used. Thus,
the scalar product of the momentum $m \mathbf{v}$ with the equation of motion (\ref{eq: motion2}) and eqs. (\ref{eq: En2}) and 
(\ref{eq: property}) yields \cite{Ko-JPA-2007}
\begin{eqnarray} \label{eq: en_rate}
\frac{d}{dt} \la E \ra &=& - \ga^{}_R \frac{1}{m} \la (\nabla S)^2 \ra + \left \la \frac{\pa V}{\pa t} \right\ra + \left \la \mathbf{r} \cdot \frac{d \mathbf{F}_r(t)}{dt} \right\ra 
\end{eqnarray}
where $ \la \pa Q / \pa t  \ra = 0 = \la \pa \ln \rho / \pa t  \ra $ has been used.

\subsection{Gaussian wave packet solution and Bohmian trajectories}

In this section, for simplicity, we restrict ourselves to the one dimensional motion. We solve eqs. (\ref{eq: con2}) and (\ref{eq: motion2}) 
by the time-dependent Gaussian ansatz (\ref{eq: rho_ansatz}) for the probability density.
This Gaussian ansatz satisfies eq. (\ref{eq: con2}) provided that the velocity field is given by
\begin{eqnarray} \label{eq: BM_vel_Gauss}
v(x, t) &=& \frac{{\dot{\si}(t)}}{\si(t)} (x-q(t)) + \dot{q}(t) 
\end{eqnarray}
from which Bohmian trajectories are easily extracted from the following expression 
\begin{eqnarray} \label{eq: BM-trajs} 
x(x^{(0)}, t) &=& q(t) + (x^{(0)} - q(0)) \frac{\si(t)}{\si(0)} 
\end{eqnarray}
where $x^{(0)}$ and $q(0)$ are the initial conditions for $x(x^{(0)}, t)$ and $q(t)$, respectively.
Eqs.  (\ref{eq: BM_vel_Gauss}) and (\ref{eq: BM-trajs}) display the standard dressing scheme where the Bohmian velocity and position are formed by the classical counterpart due to the center of the wave packet (particle property) plus a term involving its width (wave property)
\cite{NaMi-book-2017}.

Introducing now ansatz (\ref{eq: rho_ansatz}) and eq. (\ref{eq: BM_vel_Gauss}) into the equation of motion (\ref{eq: motion2}), one obtains
\begin{numcases}~
\ddot{q}(t) + \ga^{}_R \dot{q}(t) + \frac{1}{m} \frac{\pa V}{\pa x} \bigg|_{x=q(t)} = F_r(t) , \label{eq: xbar}  \\
\ddot{\si}(t) + \ga^{}_R \dot{\si}(t) - \frac{\hb^2}{4 m^2 \si(t)^3} - \frac{\hb \ga^{}_I }{2m}  \frac{1}{\si(t)} + \si(t) \frac{1}{m} \frac{\pa^2 V}{\pa x^2} \bigg|_{x=q(t)} = 0  \label{eq: delta} 
\end{numcases}
where we have used the wave packet approximation to expand the interaction potential around the classical path $q(t)$ up to second order. 
As can be seen, $q(t)$ is ruled by the classical Langevin equation of motion and $\sigma(t)$ by a generalized Pinney equation; the new term accompanying $\gamma_I$ and going as $\sigma(t)^{-1}$ provides the extension of the standard dissipative Pinney equation \cite{NaMi-book-2017,MoMi-submitted-2019}.
The Pinney equation is also known as Ermakov equation and appears, for example, in the
process of cooling down atoms in a harmonic trap \cite{Mu-PRL}. 
It is notable that the differential
equations governed by $q(t)$ and $\sigma (t)$ are not coupled each other; the width is not influenced by the random force. 
Chavanis \cite{Ch-EPJP-2017}
mentions that in cosmology  the so-called Hubble parameter which can be defined as the ratio $\dot \sigma / \sigma$, the corresponding width
also follows eq. (\ref{eq: delta}). In any case, this generalized Pinney equation has not been reported in the literature within  this context. He 
also speaks  about the quantum damped isothermal Euler equation since the imaginary part of the friction coefficient is related to a statistical potential. 
For time-independent quadratic potentials there is a soliton-like solution for (\ref{eq: delta}) where the width of the wave packet remains constant, $ \dot{\si} = 0 $,
\begin{eqnarray} \label{eq: soliton}
\ga^{}_I &=& - \frac{\hb}{2 m \si_0^2} + \frac{2 \si_0^2}{\hb} V_2 
\end{eqnarray}
where $ V_2 = d^2 V(x) / d x^2 $ is a constant for time-independent quadratic potentials.
One can rewrite this equation as a condition for the initial value of the width to have a soliton-like solution. 
Eq. (\ref{eq: soliton}) reveals that for free particles $\ga^{}_I$ must be negative in order to have a soliton-like solution. 
For such a solution quantum diffusion coefficient is the same as that of the classical mechanics \cite{MoMi-submitted-2019}.

\section{Results}

The generalized Schr\"{o}dinger equation (\ref{eq: gen_SCH2}) is going to be the starting point of the dissipative and
stochastic dynamics analyzed here for simple systems, that is, the motion of a single particle immerse in a given environment. For this
purpose,  $\gamma_R$ and $\gamma_I$ are considered as two parameters of the theory. 
As mentioned before, the Gaussian solution of this equation leads to eqs. (\ref{eq: xbar}) and (\ref{eq: delta}) responsible respectively for the center and the width of the Gaussian wave packet.
For analytical and numerical calculations of this section, we take the random force a delta-correlated Gaussian white noise with average zero,
\begin{numcases}~ 
\la F_r(t) \ra = 0 , \label{eq: white} \\
\la F_r(0) F_r(t) \ra = 2m \ga k_B T \del(t)  \label{eq: fluc-diss}
\end{numcases}
where $T$ is the bath temperature and $\del(t)$ is the Dirac delta function.

\subsection{The position-momentum uncertainty relation for free dissipative dynamics}

As is known, the expectation values of the momentum operator and its square are given by
\begin{eqnarray} 
\la \hat{p} \ra &=& -i \hb \int dx~ \psi^*(x, t) \frac{\pa}{\pa x} \psi(x, t)
= \int dx ~ \rho \frac{ \pa S }{ \pa x }
\label{eq: mom1_exp}
 \\
\la \hat{p}^2 \ra &=& \hb^2 \int dx~ \left| \frac{\pa \psi(x, t)}{\pa x}  \right|^2
= \int dx ~ \rho(x, t) \left( \frac{ \pa S }{ \pa x } \right)^2
+ \hb^2 \int dx ~ \left( \frac{ \pa |\psi| }{ \pa x } \right)^2
\label{eq: mom2_exp}
\end{eqnarray}
where in the second equation we have used the square-integrable property of the wave function.
From the Gaussian ansatz (\ref{eq: rho_ansatz}) and velocity field (\ref{eq: BM_vel_Gauss}), one obtains
\begin{eqnarray}
\la \hat{p} \ra &=& p(t) , \\
\la \hat{p}^2 \ra &=& 
p(t)^2 + m^2 \dot{\si}(t)^2 + \frac{\hb^2}{4 \si^2(t)} 
\end{eqnarray}
with $p(t) = m \dot{q}(t)$. Thus, with respect to the position uncertainty $\Delta x(t) = \si(t)$  we have that
\begin{eqnarray}
U(t) & \equiv & \Delta x(t) \Delta p(t) = 
\sqrt{ \frac{\hb^2}{4} + m^2 \si(t)^2 \dot{\si}(t)^2  }
\geq  \frac{ \hb }{ 2 }
\end{eqnarray}
for the uncertainty product where $ \Delta p (t) = \sqrt{\la \hat{ p }^2 \rangle - \langle \hat{ p } \rangle^2 } $ is the uncertainty in momentum.

\begin{figure} 
\centering
\includegraphics[width=12cm,angle=0]{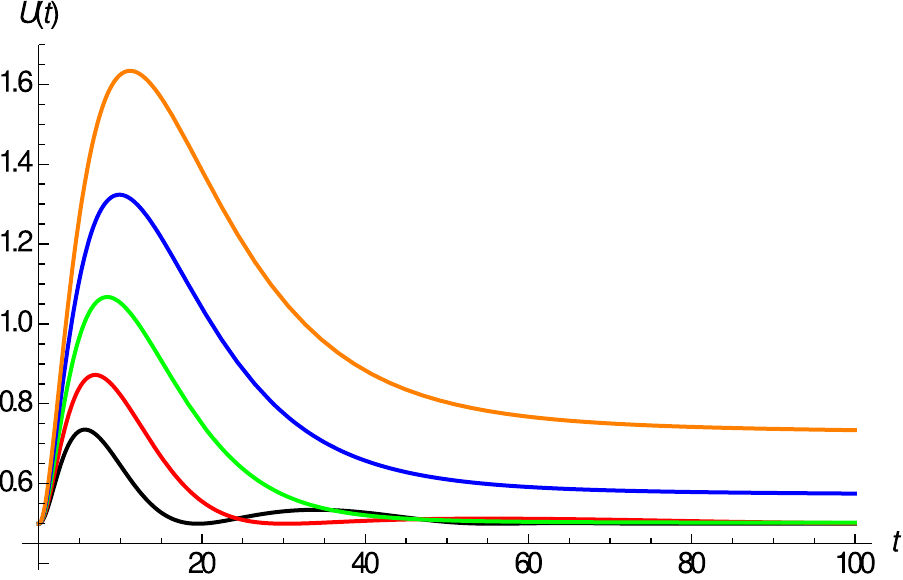}
\caption{
The uncertainty product versus time for a free dissipative dynamics: $\ga^{}_I = -0.1$ (black curve), $\ga^{}_I = -0.05$ (red curve), 
$\ga^{}_I = 0$ (green curve), $\ga^{}_I = 0.05$ (blue curve) and $\ga^{}_I = 0.1$ (orange curve). For numerical calculations we have 
used $m=1$, $\hb =1$, $\si_0=1$, $q(0)=-20$, $\dot{q}(0)=4$ and $\ga^{}_R = 0.1$.
}
\label{fig: Uncer} 
\end{figure}

In Figure \ref{fig: Uncer}, the uncertainty product is plotted for a free dissipative dynamics ($V=0$ and $F_r=0$)
and for a given $\ga^{}_R$ but different values 
of $\ga^{}_I$: $\ga^{}_I = -0.1$ (black curve), $\ga^{}_I = -0.05$ (red curve), $\ga^{}_I = 0$ (green curve), $\ga^{}_I = 0.05$ (blue curve) and 
$\ga^{}_I = 0.1$ (orange curve). For numerical calculations we have used $m=1$, $\hb =1$, $\si_0=1$, $q(0)=-20$, $\dot{q}(0)=4$ and $\ga^{}_R = 0.1$. 
One clearly sees that the uncertainty product increases first and after reaching a maximum value depending on $\ga^{}_I$, it decreases smoothly except for $\ga^{}_I=-0.1$ which another maximum is seen.
Note that for the soliton-like solution (\ref{eq: soliton}) with $\ga^{}_I=-0.5$, the uncertainty product is independent of time and has the minimum value $0.5$.
The asymptotic values observed for each case are greater than 0.5, increasing with the imaginary part of the friction coefficient.

\subsection{Diffusion coefficient for the Brownian-Bohmian motion}

In this subsection we are going to take into account the effect of thermal fluctuations in the environment  ($F_r \neq 0$). 
The simplest system is the Brownian motion of a particle subject only to a Gaussian white noise. When considering this motion in the Bohmian framework, we talk about the Brownian-Bohmian motion \cite{NaMi-book-2017}.
For free propagation ($ V = 0 $) solution of the classical Langevin equation (\ref{eq: xbar}) is given by \cite{MoMi-submitted-2019}
\begin{eqnarray*} \label{eq: xt_free_formal_sol} 
q(t) &=& q(0) + \dot{q}(0) \frac{1}{\ga} (1- e^{-\ga t} ) 
+ \frac{1}{m \ga} \int_0^t d\tau F_r(\tau) e^{-\ga ( t - \tau) } .
\end{eqnarray*}
Then from the properties (\ref{eq: white}) and (\ref{eq: fluc-diss}) of the random force and the Maxwell-Boltzmann distribution for the initial velocities, $ f_T(\dot{q}(0))  = \sqrt{ \frac{m}{2\pi k_B T} } \exp \left[ -\frac{m \dot{q}(0)^2}{2 k_B T} \right]$, one obtains
\begin{eqnarray} \label{eq: MSD_cl}
\la\la ( q(t) - q(0))^2 \ra\ra & = & 2\frac{k_B T}{m \ga^{}_R} \left( t - \frac{1-e^{-\ga^{}_R t}}{\ga^{}_R} \right)
\end{eqnarray}
for the mean squared displacement (MSD) where the double averaging implies average over the noise and  the initial velocities which are distributed according to 
the Maxwell-Boltzmann distribution function, $ \la \la \cdots \ra \ra = \int d\dot{q}(0) ~ f_T(\dot{q}(0)) ~ \la \cdots \ra $.

Eq. (\ref{eq: MSD_cl}) implies that in the diffusion regime $ t \gg \ga^{-1}_R $, MSD is 
proportional to time with a constant given by $ 2 D $ where $ D =  k_B T / m \ga_R $ is the diffusion constant (Einstein's law). A time-dependent 
diffusion coefficient $ D(t)$ can be defined as the ratio of MSD over $2t$ \cite{MoMi-submitted-2019}.
From eqs. (\ref{eq: BM-trajs}) and (\ref{eq: MSD_cl}), 
MSD $ \la\la (x(x^{(0)}, t) - x^{(0)})^2 \ra \ra $ of Bohmian stochastic trajectories averaged over initial Bohmian positions $ x^{(0)} $ according 
to the Born distribution rule gives \cite{MoMi-submitted-2019}
\begin{eqnarray} \label{eq: BM_D}
D_{\qm}(t) &=& D_{\cl}(t) + \frac{1}{2 t} ( \si(t) - \si(0) )^2 ,
\qquad D_{\cl}(t) = \frac{k_B T}{m \ga^{}_R} \left( 1 - \frac{1-e^{-\ga^{}_R t}}{\ga^{}_R t} \right)
\end{eqnarray}
for the quantum diffusion coefficient where $ D_{\cl}(t) $ is the corresponding classical quantity.
For the soliton-like solution (\ref{eq: soliton}), quantum diffusion coefficient is exactly the same as that of classical mechanics which leads to 
the same diffusion constant for both classical and quantum mechanics.

\begin{figure} 
\centering
\includegraphics[width=15cm,angle=0]{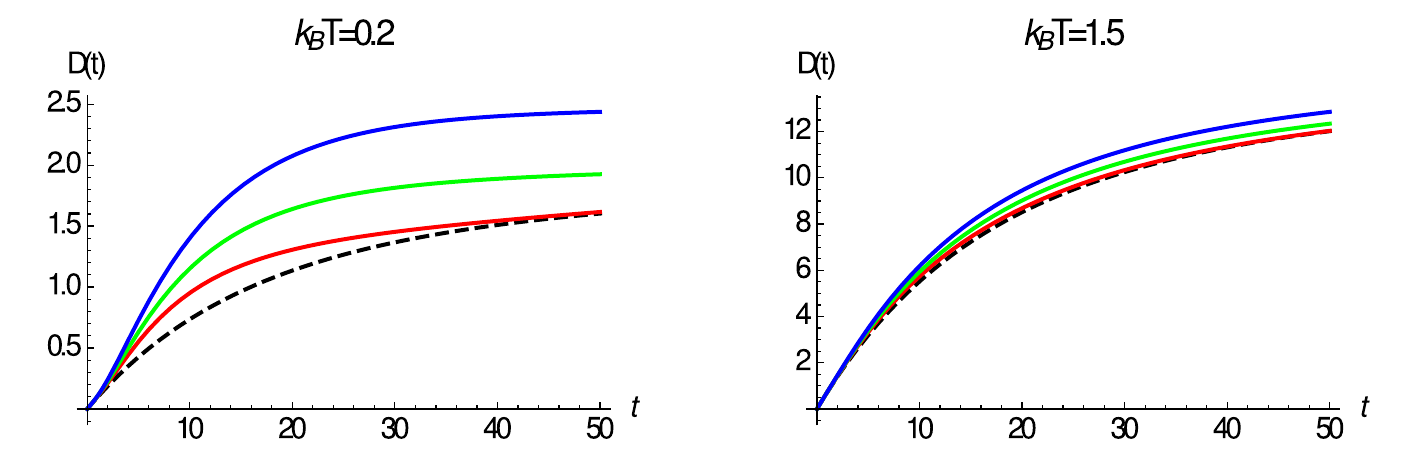}
\caption{
Classical diffusion coefficient (dashed black curves) and quantum diffusion coefficients versus time for $k_B T = 0.2$ (left panel) and 
$k_B T = 1.5$ (right panel) for Brownian-Bohmian motion for $\ga^{}_I = -0.1$ (red curves), $\ga^{}_I = 0$ (green curves) and 
$\ga^{}_I = 0.1$ (blue curves). Other parameters are the same as those of Fig. \ref{fig: Uncer}.
}
\label{fig: diff} 
\end{figure} 

In Fig. \ref{fig: diff}, the  classical (dashed black curves) and quantum diffusion coefficients (color curves) versus time are plotted for 
$k_B T = 0.2$ (left panel) and $k_B T = 1.5$ (right panel) for the Brownian-Bohmian motion with $\ga^{}_I = -0.1$ (red curves), 
$\ga^{}_I = 0$ (green curves) and $\ga^{}_I = 0.1$ (blue curves). Other parameters are the same as those of Fig. \ref{fig: Uncer}. It is clearly
seen in both panels that eah curve tends to an asymptotic value from which the corresponding diffusion constant can be extracted. As expected
from Einstein's relation, this constant is greater for higher temperatures. Even more, quantum constants are always  greater than classical ones 
due to the width contribution of the wave packet after eq. (\ref{eq: BM_D}).
It should be again emphasized here that $q(t)$ follows a classical Langevin equation with friction $\gamma_R$ and $\sigma(t)$ depends on 
$\gamma_R$ and $\gamma_I$ at the same time according to the generalized Pinney equation (\ref{eq: delta}).

\subsection{Stochastic transmission through a transient parabolic repeller. Early arrivals}

A number of interesting phenomena are seen in time-dependent quantum systems. Among these, one can mention the phenomenon of 
{\it early arrivals} \cite{HoMaMa-JPA-2012, superarrivals} which has been reported in the scattering of wave packets from time-dependent 
barriers for  isolated systems. 
Before the transmission probability reaches its stationary value, there is a time-interval where an enhancement of this probability with time 
is seen as compared to the case of free wave packet propagation.
Early arrivals refer to this early increase (relative to the free case) in the transmission probability. 
It is then very illustrative to study this effect for open quantum systems. To this end we consider stochastic transmission from a time-dependent barrier
\begin{eqnarray} \label{eq: para_rep}
V(x, t) &=& - \frac{1}{2} m \om^2 e^{-g(t-t_B)^2} x^2
\end{eqnarray}
which corresponds to the appearance of a parabolic repeller during a short time interval by choosing a Gaussian form for the time window, 
with the parameters $ t_B $ and $g$ displaying the peak time and inverse width of the window, respectively. Here, $\om$ characterizes the strength 
of the barrier. 
Let us consider a wave packet initially well localized in the left side of the barrier which is sent towards the barrier. The transmission probability 
is given by
\begin{eqnarray} \label{eq: tr_prob}
P_{\tr}(t) &=& \int_{x_d}^{\infty} dx ~ \rho(x, t)
= 
\frac{1}{2} \mbox{erfc} \left[ \frac{x_d - q(t)}{\sqrt{2} \si(t)} \right]
\end{eqnarray}
where $ x_d $ is the detector location and in the second equality we have used the Gaussian ansatz (\ref{eq: rho_ansatz}). For the pure 
dissipative dynamics, we fix $t_B$ by $ q_f(t_B)  = 0 $ where $f$ denotes
the free case i.e., the strength of the barrier is maximum when the center of the {\it free} Gaussian packet arrives at the top of the barrier. 
Thus, from the solution of eq. (\ref{eq: xbar}) for the free dissipative dynamics we have
\begin{eqnarray}
t_B &=& -\frac{1}{\ga^{}_R} \ln \left( 1 + \ga^{}_R \frac{q(0)}{\dot{q}(0)} \right)   .
\end{eqnarray}
Noting the negative sign of $q(0)$ and positive $ \dot{q}(0) $ and in order to have a positive time $ t_B $ we must impose the condition
\begin{eqnarray}
\ga^{}_R \frac{|q(0)|}{\dot{q}(0)} &<& 1    .
\end{eqnarray}
\begin{figure} 
\centering
\includegraphics[width=15cm,angle=0]{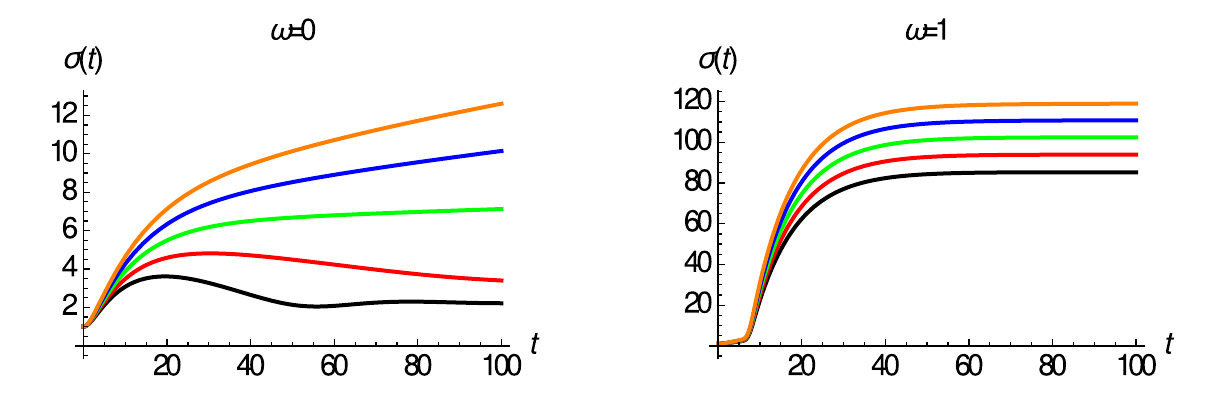}
\caption{
Width of the wave packet versus time for a dissipative dynamics in free motion, $\om= 0$ (left panel), and $\om=1$ (right panel) for different
values of the imaginary part of the friction coefficient:
$\ga^{}_I = -0.1$ (black curves), $\ga^{}_I = -0.05$ (red curves), $\ga^{}_I = 0$ (green curves), $\ga^{}_I = 0.05$ (blue curves) and 
$\ga^{}_I = 0.1$ (orange curves). For numerical calculations we have used $g_0=1$ and $x_d = 20$. Other parameters are the same as those 
used in Fig. \ref{fig: Uncer}.
}
\label{fig: width_diss} 
\end{figure}

In Figure \ref{fig: width_diss}, the width of the wave packet is plotted versus time for the free dissipative case, $\om= 0$ (left panel), and $\om=1$ 
(right panel) for different values of the imaginary part of the friction coefficient:
$\ga^{}_I = -0.1$ (black curves), $\ga^{}_I = -0.05$ (red curves), $\ga^{}_I = 0$ (green curves), $\ga^{}_I = 0.05$ (blue curves) and 
$\ga^{}_I = 0.1$ (orange curves). For numerical calculations, $g_0=1$ and $x_d = 20$ have been chosen. Other parameters are the same as those 
of Fig. \ref{fig: Uncer}. These results show that the width increases globally with the strength of the barrier and with $\gamma_I$ and its 
asymptotic value is rapidly reached for the highest strength case. Furthermore, negative values of $\gamma_I$ lead to narrower widths, being
the time dependence not so regular.
This behavior is in accord to the results of Ref. \cite{BBMy-AP-1976} where a logarithmic potential added to the usual {\it linear} 
Schr\"{o}dinger equation converts it to a non-linear equation and acts against the spreading of the wave packet.

\begin{figure} 
\centering
\includegraphics[width=12cm,angle=0]{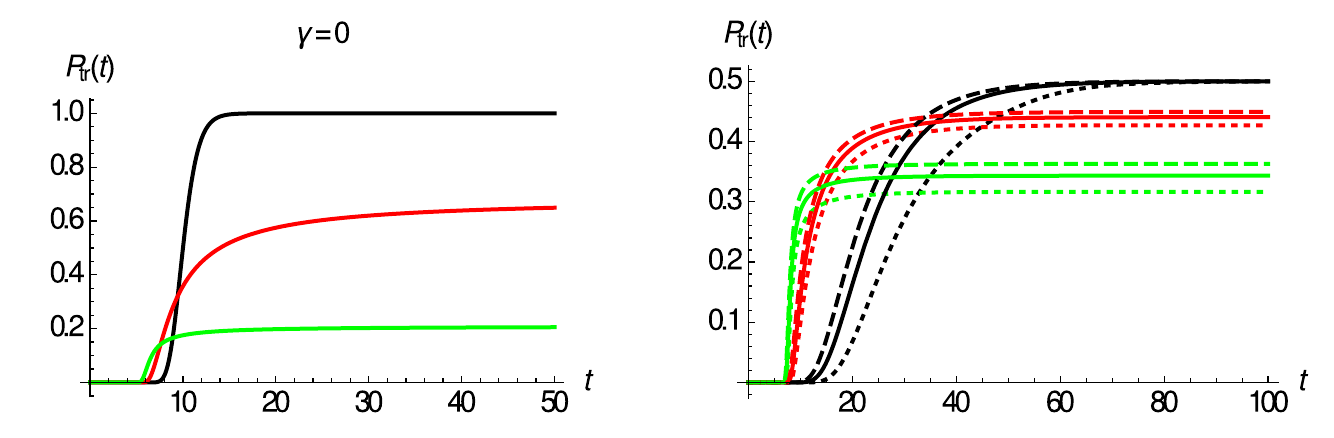}
\caption{
Time-dependent transmission probability for non-dissipative dynamics (left panel) and dissipative dynamics (right panel) with $\ga=0.1(1-i)$ 
(dotted curves), $\ga=0.1$ (solid curves) and $\ga=0.1(1+i)$ (dashed curves) for different values of barrier's strength $\om=0$ 
(black curves), $\om=0.8$ (red curves) and $\om=1.5$ (green curves). Other parameters are the same as those of 
Fig. \ref{fig: width_diss}.
}
\label{fig: Trprob} 
\end{figure}

Concerning transmission probability,  since $t_B$ and $q(t)$ are dependent on $\ga^{}_R$ and not on $\ga^{}_I$ after the behavior of 
the complementary error function, it is concluded that this probability should increase with the imaginary part of the friction coefficient when 
its real part remains constant. In Figure \ref{fig: Trprob}, the transmission probability (\ref{eq: tr_prob}) versus time is plotted 
for a non-dissipative dynamics (left panel) and different dissipative dynamics (right panel) for a fixed value of $\ga^{}_R$ but different 
values of $\ga^{}_I$ and barrier's strengths. In particular, the following values have been used: $\ga=0.1(1-i)$ 
(dotted curves), $\ga=0.1$ (solid curves) and $\ga=0.1(1+i)$ (dashed curves) for different values of barrier strengths: $\om=0$ 
(black curves), $\om=0.8$ (red curves) and $\om=1.5$ (green curves). Other parameters are the same as those used in
Fig. \ref{fig: width_diss}.
In both dynamics, the stationary value of transmission probability decreases with the barrier strength. In the dissipative case and for a given 
$\om$, transmission increases with $\ga^{}_I$ keeping $\ga^{}_R$ fixed.
Furthermore,  there is always a time interval during which the time-dependent transmission probability is higher for the interacting case 
than for the free case. This is the so-called superarrivals or early arrivals phenomenon which is seen in transmission through transient barriers \cite{HoMaMa-JPA-2012, superarrivals}. A possible application for a key generation and a procedure to speed-up entanglement between two 
qubits has been proposed elsewhere \cite{HoMaMa-JPA-2012}. 
Early arrival can be quantified by the ratio \cite{HoMaMa-JPA-2012, superarrivals}
\begin{eqnarray} \label{eq: super}
\eta_{\om} &=& \frac{I_{\om}-I_f}{I_f}
\end{eqnarray}
where 
\begin{eqnarray} \label{eq: super_I}
I_{\om} &=& \int_{t_d}^{t_c}  dt ~ P_{\tr, \om}(t) ; \qquad 
I_{f} = \int_{t_d}^{t_c}  dt ~ P_{\tr, f}(t)
\end{eqnarray}
are respectively the surface below the time-dependent transmission probability in the interacting and free cases during the time interval $\Delta t = t_c - t_d $ over which superarrival takes place. 
At time $t_d$ the curve of transmission probability for the interacting case deviates from the corresponding curve for the free case while at $t_c$ both curves cross. 
Now, we examine the magnitude of early arrival for the given value $\om=1.5$ of the barrier strength.
From the information contained in Figure \ref{fig: Trprob} and by using equation (\ref{eq: super}) Table \ref{table} is obtained.
\begin{table}
\begin{center}
 \begin{tabular}{ c || c | c | c | c | c | c |} 
 \hline
  & $~~t_d~~$ & $t_c$ & $ I_f $ & $ I_{1.5} $ & $ \eta_{_{1.5}} $ \\ [0.5ex] 
 \hline\hline
 $\ga = 0$ & 5.2 & 8.8912 & 0.0814939 & 0.377325  & 3.6301   \\ 
 \hline
 $\ga = 0.1(1+i)$ & 6.8 & 26.235 & 2.71907 & 6.28026 & 1.30971 \\
 \hline
 $\ga = 0.1$ & 6.9 & 27.74212 & 2.52066 & 6.37009 & 1.52715 \\
 \hline
 $\ga = 0.1(1-i)$ & 7 & 32.835 & 2.71218 & 7.36953 & 1.71719 \\ [1ex] 
 \hline
\end{tabular}
\caption{Early arrival characterization from the ratio defined in eq. \ref{eq: super}.\label{table}}
\end{center}
\end{table}
In this Table, we have chosen the deviation time $t_d$ in a way that transmission probability at this time is of the order of $ 10^{-4} $ for 
the interacting case while it is $ \sim 10^{-11} $ for the free propagation. At the cross time $t_c$ transmission time for both free and 
interacting case are the same to four digits decimals. 
It should be mentioned that the choice zero as the lower limit of integrals in (\ref{eq: super_I}) instead of $t_d$ has negligible effect on the above results. 
In any case, we see that the magnitude of early arrival reduces for dissipative dynamics and for such a dynamics early arrival is higher for negative $\ga^{}_I$.

\begin{figure} 
	\centering
	\includegraphics[width=10cm,angle=-90]{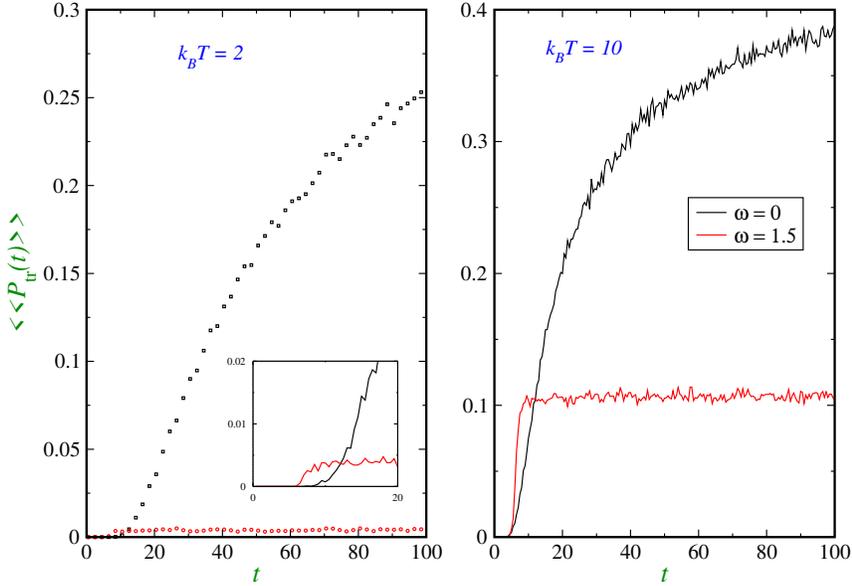}
	\caption{
		Time-dependent transmission probability in the presence of noise for $ k_B T = 2 $ (left panel) and $ k_B T = 10 $ (right panel) and for 
		$ \ga_I = 0 $ for different values of barrier's strength $\om=0$ (black curves) and $\om=1.5$ (red curves). Other parameters are the same as 
		those used in Fig. \ref{fig: width_diss}. The inset in the left panel clearly displays the early arrival phenomenon.
	}
	\label{fig: Trprob_noise} 
\end{figure}

Finally, in Figure \ref{fig: Trprob_noise}, the time-dependent transmission probability under the presence of thermal fluctuations or noise are plotted. 
The Langevin equation (\ref{eq: xbar}) is solved by using an algorithm proposed in \cite{VaCi-CPL-2006} with initial conditions 
$ q(t)|_{t=0} = q(0) $ and $ \dot{q}(t)|_{t=0} = \dot{q}(0) $ where $ \dot{q}(0) $ has a Maxwell-Boltzmann distribution. Then for the thermal transmission probability we have
\begin{eqnarray}
\la \la P_{\tr}(t) \ra \ra &=& 
\sum_{i=1}^{n_{\text{tra}}} P_{\tr, i}(t) =
\frac{1}{2}  \sum_{i=1}^{n_{\text{tra}}} \mbox{erfc} \left[ \frac{x_d - q_i(t)}{\sqrt{2} \si(t)} \right]
\end{eqnarray}
where in the second equality we have used relation (\ref{eq: tr_prob}) and $q_i(t)$ refers to the $i-$th trajectory. For the number of trajectories used to produce a Maxwell-Boltzmann distribution for initial velocities we use $ n_{\text{tra}} = 10000 $.
Two different temperatures $ k_B T = 2 $ (left panel) and $ k_B T = 10 $ (right panel) are used for $ \ga^{}_R = 0.1 $ and $ \ga^{}_I = 0 $. 
In each panel, two different values of barrier's strength $\om=0$ (black curves) and $\om=1.5$ (red curves) are showed.
Here, we have set $ t_B $ in eq. (\ref{eq: para_rep}) as $ t_B = 3 t_b $ where $ t_b = 2m\si_0^2/\hb $ is the time-scale appearing in the 
relation of freely propagating Gaussian wave packet in a non-dissipative medium.
Again, the phenomenon of early arrivals is seen in this stochastic dynamics.
One clearly sees that temperature enhances transmission probability. 
We observe similar behaviors for non-zero values of $\ga^{}_I$.

\section{Concluding remarks}

Recently, Chavanis  has proposed two generalized Schr\"{o}dinger equations for quantum dissipative systems which although both globally 
conserve probability density only one fulfills local conservation of the normalization. In both equations, friction coefficient is a complex quantity 
which the imaginary part (quantum friction) is interpreted as an effective temperature leading to a statisitcal potential when considering an ensemble
of particles. 
Within the Bohmian mechanical framework, both equations have been analyzed in terms of quantum trajectories by considering a Gaussian ansatz 
for the probability density and simple systems. In the first equation, the center of the wave packet  does not follow a classical trajectory revealing
that Ehrenfest theorem is violated.   
We also show contrary to some claims that the SCH equation violates Ehrenfest theorem. 
The SCH equation is a special case of the NM equation proposed for continuous measurements. For the Gaussian solution, we have also 
showed by a correct velocity field this equation preserves equivariance property needed for equivalence between results of Bohmian mechanics 
with those of standard quantum theory.

The second generalized equation is equivalent to the Kostin equation for real frictions while reduces to Bialynicki-Birula and Mycielski equation 
for imaginary frictions. From this equivalence, we have interpreted the imaginary part of the complex friction coefficient as a factor responsible for 
soliton-like solutions. In fact, we have observed that for a given negative $\ga^{}_I$, the width of the Gaussian wave packet remains constant in its free 
propagation. With a Gaussian ansatz for the probability density, Bohmian stochastic trajectories are again obtained.  

%

\vspace{1cm}
\noindent
{\bf Acknowledgement}
\vspace{1cm}

SVM acknowledges support from the University of Qom and SMA support from 
the Ministerio de Ciencia, Innovaci\'on y Universidades (Spain) under the
Project FIS2017-83473-C2-1-P.



%
\end{document}